\documentstyle[epsf,graphicx]{mn}

\def\etal{{et al.\ }}

\def\x2{$\chi^{2}$}

\def\asca{{\it ASCA }}
\def\rosat{{\it ROSAT }}

\def\x2{$\chi^{2}$}
\def\lunits{$\rm{erg\,s^{-1}}$}

\def\cunits{$\rm{cm^{-2}}$}

\newbox\grsign \setbox\grsign=\hbox{$>$} \newdimen\grdimen \grdimen=\ht\grsign
\newbox\simlessbox \newbox\simgreatbox \newbox\simpropbox
\setbox\simgreatbox=\hbox{\raise.5ex\hbox{$>$}\llap
     {\lower.5ex\hbox{$\sim$}}}\ht1=\grdimen\dp1=0pt
\setbox\simlessbox=\hbox{\raise.5ex\hbox{$<$}\llap
     {\lower.5ex\hbox{$\sim$}}}\ht2=\grdimen\dp2=0pt
\setbox\simpropbox=\hbox{\raise.5ex\hbox{$\propto$}\llap
     {\lower.5ex\hbox{$\sim$}}}\ht2=\grdimen\dp2=0pt

\begin{document}

\title[ \rosat observations of UGC4305] {\rosat observations of the dwarf starforming galaxy Holmberg
II (UGC 4305)}

\author[A.L. Zezas, I. Georgantopoulos  and M.J. Ward]
       {A.L. Zezas$^{1}$, I. Georgantopoulos$^{2}$ and M.J. Ward$^{1}$ \\
 Department of Physics and Astronomy, University of Leicester, 
Leicester, LE1 7RH \\
 National Observatory of Athens, Lofos Koufou, Palaia Penteli, 
15236, Athens, Greece } 

\maketitle

\label{firstpage}

\begin{abstract}
We present \rosat PSPC and HRI observations of the dwarf 
irregular galaxy Holmberg II (UGC4305). This is one of the 
most luminous dwarf galaxies ($L_x\sim 10^{40}$ \lunits) 
detected in the \rosat All-Sky Survey. The X-ray emission 
comes from a single unresolved point source, coincident with 
a large HII region which emits intense radio emission. 
The source is variable on both year and day timescales, 
clearly favouring  accretion into a compact object rather than 
a supernova remnant or a superbubble interpretation for the origin 
of the X-ray emission. 
However,  its X-ray spectrum  is well-fit by a 
a Raymond-Smith spectrum with kT$\sim$0.8 keV,  lower than the 
temperature of X-ray binaries in nearby spiral galaxies.  

\end{abstract}


\section{INTRODUCTION}

Dwarf starforming galaxies have been extensively studied in X-rays
during the last few years. Their X-ray luminosities range 
from around $10^{38}$ \lunits (eg NGC1569; Della Ceca et al. 1996) up to 
few times $10^{40}$ \lunits (eg Mrk 33; Stevens \& Strickland 1998) 
 more than the whole X-ray luminosity of the Local Group.
Many objects appear to be extended in soft X-rays (0.5-2 keV): 
 eg NGC1569 (Heckman et al. 1995),  NGC4449  (Della Ceca \etal 1997), 
 He 2-10 (Hensler, Dickow \& Junkes 1997),  NGC1705 (Hensler et al. 1998),
IC2574  (Walter et al. 1998). This gave support to a 'superwind' 
 model for the origin of the X-ray emission.
 According to this model the numerous supernovae 
 in the starburst drive a galactic scale outflow. 
 Such superwinds are  observed in other star-forming 
 galaxies such as M82 (Strickland, Ponman \& Stevens 1997)
 and NGC253 (Fabbiano and Trinchieri 1984). 
 However, in the case of dwarf galaxies where the gravitational 
potential is low the effects of such superwind can 
be catastrophic as the hot gas may eventually escape from the 
galaxy ceasing the star formation. 
 The spectrum of these regions is soft (kT$\sim$0.8 keV)
 (eg Stevens \& Strickland 1998) although much softer values 
have also been reported (eg NGC1705, Hensler et al. 1997).  
 Even in cases where the emission does not appear to 
 be extended the X-ray emission is usually attributed to a 
 superbubble (eg NGC5408, Fabian \& Ward 1993; Mrk 33, Stevens \& Strickland 
 1998). In one case so far,  (IC 10, Brandt et al. 1997)
 the X-ray emission, which remains unresolved by the \rosat HRI, 
 probably originates in an X-ray binary, proving  
 that there are  multiple origins for the origin of the X-ray emission
 in dwarf star-forming galaxies. 

 In contrast to the situation for the soft X-rays, hard X-ray observations of dwarf star-forming 
galaxies are scarce. 
 Della Ceca et al. (1996) and Della Ceca, Griffiths \& 
 Heckman (1997) observed NGC1569 and NGC4449 respectively with \asca.  
Their X-ray spectrum appears to be complex. It can be  represented by 
at least two thermal components: the soft with 
 kT$\sim$0.8 keV, probably originating from the superwind, 
 while the harder component (kT$\sim$4 keV) which is spatially 
 unresolved by the \asca SIS, has an unknown origin. 
 The \asca observations clearly demonstrate 
 the complexity of the X-ray emission mechanisms 
 in dwarf star-forming galaxies again emphasizing that the 
 origin of the X-ray  emission in these objects is still an open question.

\subsection{  Holmberg II  (UGC 4305)}

 Holmberg II is one of the most luminous dwarf irregular galaxy in the sample
of Moran \etal (1996), with an X-ray luminosity of $\sim 10^{40}
\rm{erg\,s^{-1}}$ in the 0.5-2.0 keV band. The above sample 
  comes from the 
cross-correlation of the IRAS Point Source catalogue with the
 the \rosat All-Sky Survey (RASS). 
 The Moran et al. sample contains mostly AGN but also  
  luminous starburst galaxies. 
  Holmberg II is also one of the  most X-ray luminous dwarf starburst 
for its mass. The ratio of $L_{X}/M_{gal}$ is about
$4\times10^{30}\rm{erg\,s^{-1}\,M^{-1}_{\odot}}$, almost an
order of magnitude higher than that of other dwarf starbursts 
 (eg Stevens \& Strickland 1998,
 Hensler 1997).  

H$\alpha$ observations (Hodge et al. 1994) 
clearly demonstrate that there is intense star-forming 
activity in this galaxy. It is composed of many HII regions 
 (their size ranging from 96 pc up to 525 pc) which appear as
bright knots in optical images. 
The star-formation rate per unit area in Holmberg II  is found to be
$1.32\times10^{-3}\rm{M_{\odot}\,yr^{-1}\,kpc^{-1}}$ (Hunter \etal
1998). Most of the HII regions are
coincident with `holes' found in the surface density of atomic hydrogen with
VLA observations at 21cm (Puche \etal, 1992). This suggests that the
HII regions excavate the interstellar medium of the  galaxy
forming these 'holes'.
 VLA radio continuum observations  (Tongue and Westpfahl
1995) again show that most of the bright HII regions 
emit intense radio emission.   
Some regions have typical (non-thermal) supernova remnant spectrum 
while others display a thermal bremsstrahlung spectrum. 

The powerful X-ray luminosity, $L_x\sim 10^{40}$ \lunits, 
in combination with the 
proximity of Holmberg II (3.2 Mpc), make it an ideal 
case for the study of the X-ray emission mechanisms 
in dwarf star-forming galaxies. In this paper, we report the imaging
  (section 3), timing  (section 4) as well as 
  spectral  analysis (section 5) results of Holmberg-II based on four \rosat 
 PSPC and HRI observations. 

\section{OBSERVATIONS AND DATA REDUCTION}

\subsection{  The {\sl ROSAT\/}~PSPC Observations}

Holmberg II has been observed on three occasions with the Position
Sensitive Proportional Counter (PSPC, Pfefferman \etal 1987) on board
\rosat (Tr\"{u}mper \etal 1984). All the  data 
 are now publically available and were retrieved from the 
 LEDAS archive in Leicester. The details of the observations are
given in table 1. 

\begin{table}
\begin{center}
\caption{Observations log.}
\begin{tabular}{ccccc}
\hline
Instrument & Date Observation started &  Net Exposure
time \\
\hline
 PSPC & 14-4-1992 &  2827  \\
 PSPC & 29-9-1992  & 8223 \\ 
 PSPC & 14-3-1993 &  5449  \\
 HRI & 17-10-1994 &  7863 \\
\hline
\end{tabular}
\end{center}
\end{table}

 For the reduction of the two datasets we have followed  the
standard procedure, using the ASTERIX package. We excluded data with
Master Veto rate higher than 170 counts per second. This gives a net on-source
exposure time of 16 ksec in total.  Then we extracted a PSPC spectral
image cube. In the spectral fits we
have excluded all the channels below 10 
and above 201 due to the  low effective area
of the PSPC at these energies as well as to its large uncertainties.  
In order to obtain the spectrum we have extracted data from a circular
region of $\sim2.0\arcmin$ radius. The background
was estimated from an annular region between radii of 
 $15\arcmin$ and $8.8\arcmin$
from the centroid, after exclusion of the discrete sources found
with the PSS algorithm (Allen 1992) down to the $4.5\sigma$ level.

\subsection{  The {\sl ROSAT\/}~HRI Observations}

Holmberg II has also been observed with the High Resolution Imager (David
\etal 1997) on board \rosat. The FWHM of the Point Spread Function of
the XRT+HRI assembly is $\sim5\arcsec$.
Again for the reduction of the data we have used the ASTERIX
package. In the screening process we have rejected all the data with
aspect error greater than 2.
  
 \section{ SPATIAL ANALYSIS}

In order to study the spatial distribution of the X-ray emission, 
we have extracted PSPC and HRI images with pixel size of 5.0 and
1.5 arcseconds respectively.
 Figure 1 shows the HRI map overlaid on an O-band PASS image
retrieved from the Digitised Sky Survey database located at Leicester. 
 The HRI  map was created from the
original $1.5\arcsec$ pixel image after smoothing with a gaussian
 ($3.5\arcsec$ FWHM). The contours correspond to levels of 0.09,
0.13, 0.18, 0.22, 0.44, 1.11, 2.22, 6.67, 8.0
$\rm{counts\,arcsec^{-2}}$. 
As there are no other X-ray sources in the HRI field, 
the registration of the X-ray contour on to
the POSS image was achieved by assuming that errors in the pointing accuracy
and the aspect solution of the HRI are negligible.
 In reality, there is a scatter of $\sim6 \arcsec$ in the difference 
 between the HRI and optical positions of SIMBAD sources,  
 probably originating in residual star-tracker errors (Briel et al. 1997).    
As a check of the pointing accuracy we have compared the coordinates of
the centroids of the point source from the different PSPC and HRI pointings. We
found that the coordinates were the same to within $3.8\arcsec$ 
apart from  the shortest PSPC exposure where the distance between the
centroids was $\sim10\arcsec$.

\begin{figure*}
\rotatebox{270}{\includegraphics[height=12.0cm]{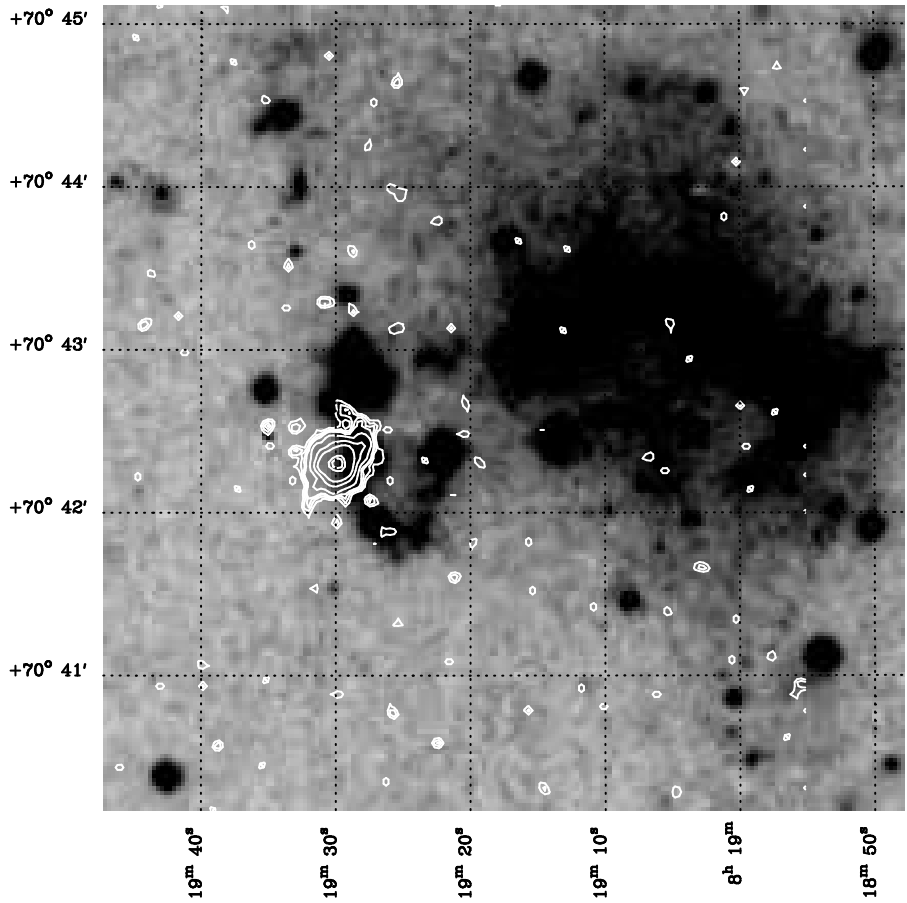}}
\caption{Contours from the HRI observation  overlaid on a
POSS image of the galaxy. The contours correspond to levels of 
0.09, 0.13, 0.18, 0.22, 0.44, 1.11, 2.22, 6.67, 8.0
$\rm{cnts\,arcsec^{-2}}$. The epoch of the coordinates is J2000.0, and
the positional error of the X-ray coordinates with respect to the
optical is $\rm{\sim6\arcsec}$.}
\end{figure*} 

 The most striking result is that the X-ray image shows only 
 one source. Comparing the radial profile of the source
with the radial profile of a point source (in this case  the star
AR-Lac, see figure 2) we  see that it is slightly extended.
 Actually, the source appears elongated in the South East - North West
direction.  However, the satellite wobbles 
in order to smooth the  efficiency variations on the microchannel plates 
 (see Briel et al. 1997). The wobble is not always appropriately 
taken into account in the aspect solution and thus 
some residual extent is possible.   
 In order to check this possibility, we extracted the HRI image in 
detector coordinates. It appears that the elongation is 
along the wobble direction and therefore we conclude that 
most probably our source is  unresolved by the HRI. 

\begin{figure*}
\rotatebox{270}{\includegraphics[height=10.5cm]{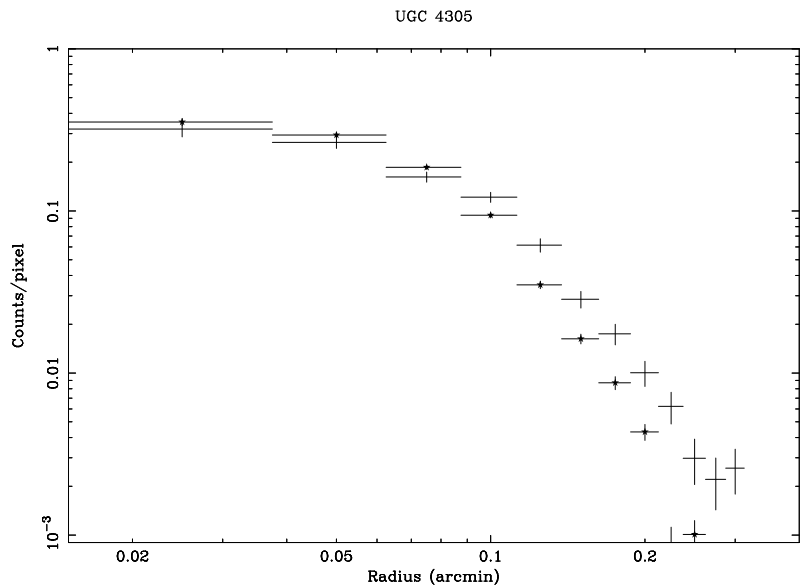}}
\caption{The HRI radial profile of the off-nuclear source in Holmberg II
(crosses) compared with the radial profile of AR-Lac (stars).}
\end{figure*}

 In order to search for low surface brightness extended X-ray
emission, we have smoothed the PSPC image using a
$1.5\arcsec$ two-dimensional gaussian. The PSPC has the advantage of
having a low
internal background and thus it can detect large-scale, low surface
brightness structures like tenuous hot gas. Again there is just one
point source in the field, as  confirmed after comparing its radial
profile with  the radial profile of Mrk 509 which we used as
a model point source (Hasinger \etal 1995).  
The X-ray source is coincident with one of the most luminous  HII regions
of Holmberg II ($\rm{L_{H\alpha}=3\times10^{38}\,erg\,s^{-1}}$). 
 Its diameter is $0.5\arcsec$, which at the galaxy's  distance 
 (3.2 Mpc) corresponds to 352pc (Hodge \etal, 1994).
Using the $logN-logS$ relation in the soft X-ray band
(0.5-2.0keV) from  Georgantopoulos \etal (1996),
we expect  $0.07 \rm{~sources\,deg^{-2}}$ to be brighter than
$\sim 10^{-12}\rm{erg\,s^{-1}\,cm^{-2}}$, 
 which is the soft X-ray flux of this galaxy.
 In the area covered by the HII 
region we expect to find $10^{-5}$ X-ray sources brighter than
$10^{-12}\rm{erg\,s^{-1}\,cm^{-2}}$ by chance, giving us confidence that 
a confusing foreground or background source is  improbable.

\section{VARIABILITY}

 In order to further investigate the nature of the X-ray source,  we
have constructed a long-term light curve from all five observations of
Holmberg II (figure 3). We have used the background subtracted count
rates and assumed a power-law model with $\Gamma=2.7$
and absorbing column density
$\rm{N_{H}=1.2\times10^{21}\,cm^{-2}}$ as found from our spectral
fits (see table 2 below). 
 The points are in chronological order; the first point
corresponds to the RASS  while the last 
to the HRI detection. All errors plotted correspond to the 
$1\sigma$ level. The errors on the PSPC observations 
  are based on counting statistics only. In contrast, the error 
 on the HRI observation includes the $1\sigma$ uncertainty 
 on the absorbing column density (see section below). 
 This is necessary in order to compare PSPC and HRI fluxes  as the 
  derived HRI flux is very sensitive on the 
 assumed column density owing to the soft energy response of the HRI.
 From this figure it is clear
that this source is variable by approximately  a factor of two.

\begin{figure}
\rotatebox{270}{\includegraphics[height=6.5cm]{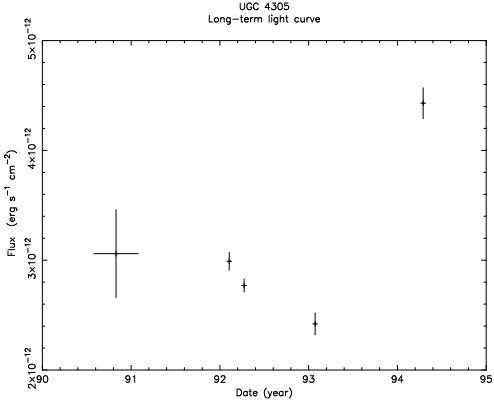}}
\caption{The long-term light curve from all the \rosat observations of
Holmberg II. The first point corresponds to the RASS 
while the last point to the HRI observation. All errors are $1\sigma$.}
\end{figure} 

 Next we extracted light curves from the four pointed 
 observations in order to check for short term variability.
 We have
extracted the background subtracted source light-curves using the
FTOOLS package. The light curves were created by 
 accumulating photons in 800 seconds bins   
in order to increase the signal to noise
ratio and smooth the effect of wobble.   
All four light-curves are presented in figure 4. 
We clearly detect short term variability, but without any obvious periodicity.
Using a \x2 test in order to check the non-variability hypothesis we
find reduced  \x2 of 41.3/11, 74.5/17, 9.4/5 and 35.1/14 for the  three
PSPC and the one HRI dataset respectively. The null hypothesis
probability is then $2\times10^{-5}, 4\times10^{-9}, 0.09$  and
0.001 for each dataset respectively.    

\begin{figure}
\includegraphics[height=10.0cm]{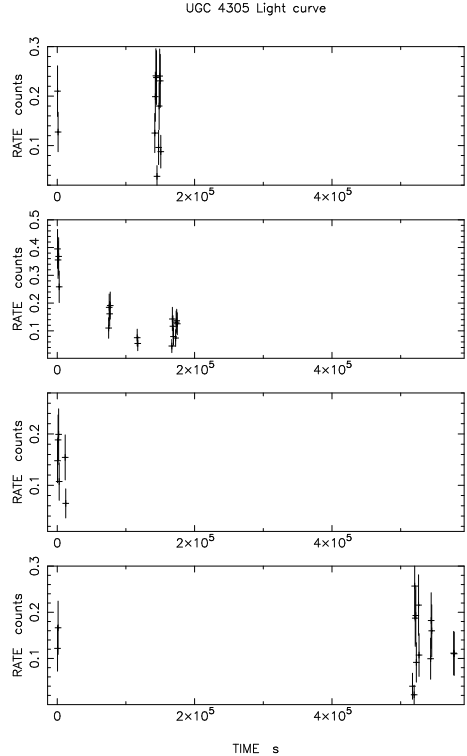}
\caption{The light curves for the four pointed observations 
 of Holmberg II. The first
three panels correspond to the three PSPC observations in
chronological order and the last panel corresponds to the HRI observation.}
\end{figure}

\section{ SPECTRAL ANALYSIS}

 For the fitting of the two PSPC spectra we used the XSPEC
package, after binning-up the spectra in order to obtain at least 20
counts in each bin. We have fitted the three spectra 
from the pointed PSPC observations simultaneously 
 (the normalizations were allowed to vary freely 
between the observations) using 
various models. The results are presented in table 2 . All the
single model fits with  absorbing column density fixed at the Galactic
value, $3.42\times10^{20}\rm{cm^{-2}}$,  (Stark \etal 1992)  
were rejected  at above the 99 per cent  confidence level.

\begin{table*}
\caption{The spectral fitting results of Holmberg II.}
\begin{tabular}{cccccc}
\hline
 Parameter  & Power-law & Single Temperature & Single Temperature &
Power-law & Raymond Smith \\
 & & Raymond Smith & Raymond Smith & Raymond Smith & Raymond Smith \\ 
\hline 
kT (KeV)  & & $3.71_{-0.68}^{+1.16}$ & $0.84_{-0.09}^{+0.12}$ &
$0.83^{+0.31}_{-0.21}$ & $2.25^{+1.54}_{-0.53}$\\
  & & & & & $0.25_{0.03}^{0.04}$  \\
 $\Gamma$  & $2.68^{+0.17}_{-0.13}$  & & &  $2.59_{-0.26}^{+0.29}$\\
 Z $\times10^{-3}$ & & & $3.36^{+1.25}_{-3.36}$ &  \\
 $N_{H} (10^{21} \rm{cm^{-2}})$  & $1.2^{+0.1}_{-0.2}$  &
 $0.47^{+0.04}_{-0.03}$ & $0.85^{+0.1}_{-0.1} $ &
 $1.1^{+0.1}_{-0.2}$ & $0.59^{+0.1}_{-0.6}$  \\ 
 $\chi^{2}$ / d.o.f. & 137.5/129 & 312.9/129 & 133.4/128 &
 131.5/127 & 141.3/127 \\
\hline
\end{tabular}

\end{table*}

 However, a power-law fit with free absorbing column density gave a
good fit with a very steep slope ($\Gamma=2.68_{-0.13}^{+0.17}$) and a
 high column density of
$12.0^{+1.0}_{-2.0}\times10^{20}\rm{cm^{-2}}$. An even lower reduced
\x2 was achieved with a Raymond-Smith thermal plasma with free
abundances and absorption. However, when compared to the  power-law model,
 this improvement is statistically significant at less than the 
  90 per cent confidence level. 
 The best-fit power law model along with
the residuals are  presented in figure 5. 
We have also tried double component models of Raymond Smith thermal
plasma (R-S) combined with another R-S or a power-law model. These  
give the same or slightly better reduced \x2 compared to the 
single component models.
 The addition of a power-law component to the 
R-S (free abundance) spectrum is only marginally statistically
significant at a level of confidence less than  $90$ per cent.

\begin{figure*}
\rotatebox{270}{\includegraphics[height=11.5cm]{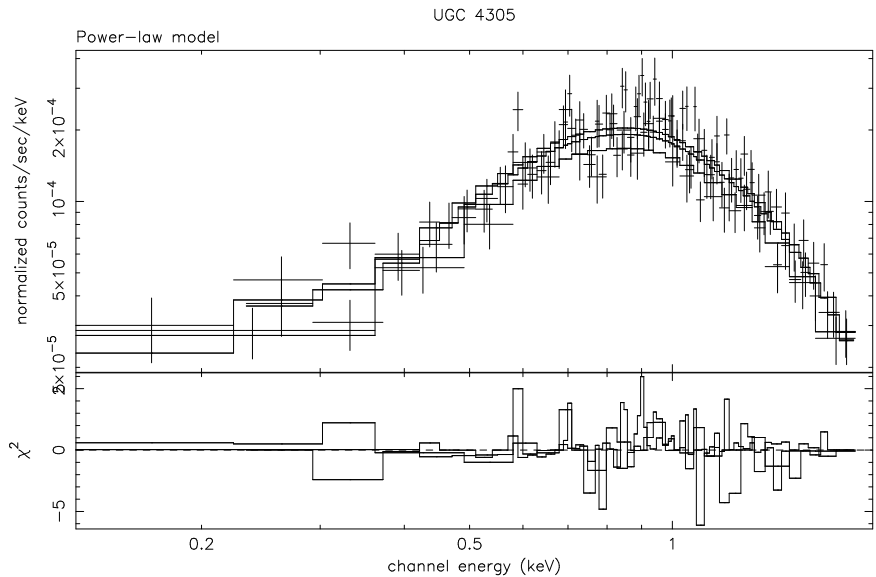}}
\caption{The best fit power-law of the PSPC X-ray spectrum of Holmberg II. 
The top panel
shows the data with the model and the bottom panel shows the
$\Delta\chi^{2}$ for the three datasets.}
\end{figure*}

\section{DISCUSSION}

The most striking result is that the X-ray emission originates from 
a single point source. Therefore, it cannot be emission from a superbubble
as in the case of NGC1569 and NGC4449. The X-ray emission of 
these two galaxies is clearly extended although these 
are situated at distances 2-5 Mpc comparable to Holmberg II. 
The X-ray emission of Holmberg II is  reminiscent of the 
point-like X-ray source in the nearby star-forming galaxy IC10 
(Brandt et al. 1997).
According to Brandt et al. the X-ray emission ($L_x\sim 4 \times 10^{38}$
 \lunits) arises from an X-ray binary. 
The unabsorbed luminosity of the X-ray source in Holmberg II 
is $3\times 10^{40}$ \lunits ,
well above the luminosities of sources detected in our Galaxy 
 and comparable with the most luminous off-nuclear sources 
detected in nearby galaxies: for example 
Marston et al. (1995) and Ehle et al. 
 (1995) report sources with luminosities of few times $10^{39}$ \lunits 
in the spiral arms of M51. 
Using \asca Reynolds et al. (1997) detect an off-nuclear X-ray 
 point source with $L_x\sim 10^{39}$ \lunits 
in the nearby spiral galaxy Dwingeloo-1, and Makishima 
et al. (1994) report a luminous  off-nuclear source in IC342,
 detected again using \asca. 
Point-like sources have also been observed in NGC 4961 and NGC 5408
(Fourniol \etal 1996, Fabian and Ward, 1993) with PSPC, having
luminosities $\rm{\sim10^{40}}$ \lunits.
  These highly luminous   
 X-ray sources  are believed to be associated with  
 supernova remnants, superbubbles or  X-ray binaries
 (Fabbiano 1995). 
Single supernovae in a dense environment 
can reach high luminosities. Fabian \& Terlevich (1996) 
report the detection of SN1988Z with a luminosity of 
$L_x\approx 10^{41}$ \lunits. 
Our source is spatially coincident with   
one of the largest HII regions in Holmberg II, a region which is also
a strong source of radio emission (Tongue \& Westpfahl 1995). 
The spectral index of the radio emission is typical of supernovae remnants. 
However, the presence of significant long-term (years)
and short term variability (days) in X-rays rules out the 
possibility that the bulk of the X-ray emission comes from 
diffuse hot gas either in a supernova or in a superbubble. 
Hence, most probably our source is associated with an 
accreting compact object. 

The X-ray spectrum of this source adds further clues 
to the origin of the X-ray emission. 
The spectrum is well-fit either by a single power-law 
with an absorption ($N_H\sim 1\times 10^{21}$ \cunits) 
well above the Galactic values ($N_H\sim 4 \times 10^{20}$ \cunits) 
or a Raymond-Smith spectrum with relatively low 
temperature (kT$\sim$ 0.8 keV), very low metallicity and
 absorption again in excess of the Galactic. 
The spectrum of the source is much harder than the typical \rosat
spectra of supernova remnants in nearby spiral galaxies, which have
temperatures of $\sim0.36$keV with a dispersion of 0.2keV (Read \etal
1997).   
On the other hand, the temperature is softer than that of 
the luminous sources detected in M51, kT$\sim$1.3-1.7keV (Marston et al. 
 1995) and of X-ray binary sources 
detected in other nearby spiral galaxies, kT$\approx$1.8keV with 
 a dispersion of 0.13 keV (Read et al. 1997). 
Interestingly, the spectrum of Holmberg II is very similar to that 
 observed in nearby Wolf-Rayet galaxies (young star-forming galaxies)
 by Stevens \& Strickland (1998). Their X-ray spectra  
have kT$\sim$0.5-1 keV, with luminosities ranging from few times 
$10^{38}$ up to $10^{41}$ \lunits while the 
metallicities of the X-ray gas are very low, typically  $Z=0.01$.
Most of these galaxies have the X-ray emission unresolved 
by the \rosat PSPC observations. According to Stevens 
\& Strickland (1998) the emission originates in 
a superbubble. As there are no timing 
observations of these galaxies, we consider it possible that 
a large fraction of the X-ray emission, at least 
in the cases of the compact dwarf galaxies,  
may originate from  the same process
as in Holmberg II. 
 As Strickland \& Stevens (1997) point out, 
  the ultrasoft components of some black hole 
candidates have roughly similar spectral characteristics (Inoue 
 1991) to  Wolf-Rayet galaxies and consequently  to Holmberg II. 
The same result is also found for galaxies in the sample of
Fourniol \etal (1996), but they cannot distinguish between a binary
or a hot gas origin of the X-ray emission in the absence of timing
data.
Assuming that the putative binary accretes at its Eddington limit,
 the mass of the central object must be
$\sim200\rm{M_{\odot}}$, 
 well in excess of the mass limit for a black hole formed by the 
 collapse of a normal star.
It is difficult to envisage how these high mass off-nuclear 
black holes formed. However, there are several ways to reduce the 
 required mass.    
 Firstly, accretion at a rate in excess of the Eddington limit:
  strong magnetic fields may channel the 
mass onto the accreting object and thus may reduce the mass 
by a factor of a few. Secondly, the emission may be anisotropic
as proposed by Reynolds et al. (1997) for Dwingeloo-X1: 
 one candidate class of objects could be the transient X-ray sources 
 with radio jets displaying superluminal motion 
 (eg GRS1915+105). 
Finally, the low metallicities derived  could 
result in the increase in the X-ray luminosity 
of an X-ray binary by as much as an order of magnitude eg. 
 van Paradijs \& McClintock (1995).

\section{CONCLUSIONS}
We have used  \rosat PSPC and HRI observations to investigate the 
X-ray properties of the nearby (3.2 Mpc) dwarf irregular galaxy Holmberg II. 
 This is  one of the most X-ray luminous 
 (unabsorbed $L_x\approx 3\times 10^{40}$ \lunits)
dwarf galaxies in the local Universe. 
Our main result is that the X-ray emission is unresolved by the HRI 
 with our X-ray source being one of the brightest off-nuclear sources 
 in nearby galaxies.  
Therefore, the X-ray emission of Holmberg II does not 
appear similar to other irregular galaxies like  NGC1569 and NGC4449,
in which the soft X-ray emission has been resolved and is believed 
to come from large superbubbles of hot gas. Instead  
the X-ray morphology is more similar to that of the small 
irregular galaxy IC 10 where the emission comes from a single point source, 
albeit with much lower luminosity. 
Our source shows strong variability (by about a factor of two) on both 
long (year) and short (days) timescales. 
The variability together with the absence of spatial extent  
clearly favour a X-ray binary scenario for the origin 
of the X-ray emission. However, the X-ray spectrum also requires interpretation. 
 The data can be well fit by a 
thermal spectrum with very low metallicity and a 
temperature of $\sim$0.8 keV, somewhat lower than the temperature 
of known X-ray binaries in nearby spiral galaxies. This 
discrepancy can be  alleviated if the emission comes 
from a black hole candidate which exhibit
similar soft spectra, or if the X-ray emission is contaminated by 
thermal emission from hot gas. 
Our analysis demonstrates the diversity of the 
X-ray emission mechanisms in dwarf galaxies.
Future high spatial resolution  and high energy 
observations with AXAF and XMM are necessary in order to 
unravel the complex X-ray properties of these objects. 

\section{Acknowledgments}
 This research has made use of data obtained through 
 the LEDAS online service, 
 provided by the University of Leicester.

\end{document}